# Biological applications of ferroelectric materials

A. Blázquez-Castro,[a)] A. García-Cabañes, and M. Carrascosa

Dept. Física de Materiales, Universidad Autónoma de Madrid, Madrid 28049, Spain

[a)]Author to whom correspondence should be addressed: alfonso.blazquez@uam.es / singlet763@gmail.com

ABSTRACT

The study and applications of ferroelectric materials in the biomedical and biotechnological fields is a novel and very promising scientific area that spans roughly one decade. However, some groups have already provided experimental proof of very interesting biological modulation when living systems are exposed to different ferroelectrics and excitation mechanisms. These materials should offer several advantages in the field of bioelectricity, such as no need of an external electric power source or circuits, scalable size of the electroactive regions, flexible and reconfigurable "virtual electrodes", or fully proved biocompatibility. In this focused review we provide the underlying physics of ferroelectric activity and a recount of the research reports already published, along with some tentative biophysical mechanisms that can explain the observed results. More specifically, we focused on the biological actions of domain ferroelectrics, and ferroelectrics excited by the bulk photovoltaic effect or the pyroelectric effect. It is our goal to provide a comprehensive account of the published material so far, and to set the stage for a vigorous expansion of the field, with envisioned applications that span from cell biology and signaling to cell and tissue regeneration, antitumoral action, or cell bioengineering to name a few.

## TABLE OF CONTENTS

# I. INTRODUCTION

Electric fields offer the possibility to move, trap and manipulate biological samples with unprecedented accuracy and reliability. Bio-object displacement comes about as the result of electrophoresis (EP), dielectrophoresis (DEP), or both. EP is the movement of charged objects under an electric field.[1] DEP is the movement of neutral objects under an inhomogeneous electric field.[2] Manipulation refers here to the outcome of changing the external and/or internal biological charge distribution that is always present in living systems and which, in fact, determines to a certain extent the "living" quality of the system. External electric fields can initiate and mimic a variety of physiological process, like the action potential in neurons or the contraction of myocytes/pacing of the heart activity. Depending on several parameters (intensity, duration, rate of change, etc.) electric fields can also induce a range of non-physiological process. For example, it is widespread the use of electric fields to eliminate tumor cells directly (electrotherapy)[3] or indirectly by enhancing antitumoral drug uptake (electrochemotherapy),[4] to promote cell uptake of compounds to which the cell is usually non-permeant (electroporation),[5] or to impose a preferred direction in cell migration (galvanotaxis).[6]

For many decades one of the main obstacles for the biological use of electricity has been the need for electrodes in order to produce the electric field. Electrodes are in general large and bulky in relation to many biological objects and, for the most, are made of metal which is not necessarily the most convenient or innocuous in a biological setup. Therefore, there has been an historical trend to reduce the physical dimensions of the electrodes, to find more biologically-friendly materials for their composition, or to design new electrode approaches that do not rely on metallic structures to allow a current to flow. In this sense, the recent years have seen many research breakthroughs in the area of alternative electrodes. A very successful approach has been that of optoelectronic electrodes, or "electrode-like" regions, which are bring about on a material as a consequence of its interaction with light. For example, the literature now abounds with reports on the generation of the so called "virtual electrodes" which refers, in fact, to the local increase in photoconductivity of a suitable substrate (usually a semiconductor) when it is conveniently illuminated.[7-11] As innovative as these approaches are, they still share some drawbacks with conventional electrodes: there remains the need for some form of external closed electrical circuit to allow current to flow, and critical from our point of view, an external electric source is mandatory for the "virtual electrode" generation.

Quite recently the possibility has been advanced to make use of ferroelectric materials to create "virtual electrodes" *ad hoc* in biological experiments.[12,13] This means that no external circuit is essential for the system to work and, most importantly, there is no need at all of an external electric source. This is related, one way or another, to the spontaneous and permanent electric polarization featured by this class of materials (see Section II below). Thanks to this feature of ferroelectrics, it is possible to harness their electric potential, under the correct conditions, to generate electric fields without an external electric source or circuitry. Three are the physical methods or approaches that will be discussed here in relation to the induction of electric fields for biological research: the use of domain structures, the bulk photovoltaic effect, and the pyroelectric effect. These approaches will be introduced in the next section and further elaborated in detail along the text in regards to biological applications. We are purposefully excluding piezoelectricity, a very broad ferroelectric application area, from this review due to several reasons. On the one hand, there are several recent reviews on this topic in relation to biological uses.[14-16] On the other hand, we want to present alternative experimental avenues to the research community to exploit the electrical activity due to ferroelectric excitation (domain structures, photovoltaic effect and pyroelectricity), which surely are not as widely known as piezoelectricity for most researchers.

It is our goal with this focused review to introduce the scientific community in general, and the applied physics and biophysics communities in particular, to the recent advancements on the use of ferroelectric materials for biological research. In this sense, Section II presents a brief and simple description of the underlying physics involved in the biological applications reported in Section III, particularly oriented to readers unfamiliar with ferroelectrics. Additionally, Section IV will be committed to present a number of tentative biophysical and biological processes that can help to explain the results published so far. We strive to introduce the topic to a larger scientific audience and set the stage for a rigorous and systematic analysis of the results obtained and the mechanisms at work.

## II. UNDERLYING PHYSICS

A ferroelectric material is a dielectric medium that presents spontaneous electric polarization $P_s$ along one crystallographic axis, the ferroelectric or polar axis. This spontaneous polarization appears without any external applied field as a consequence of the structure of the crystal lattice: positive crystal charges are slightly shifted with regard to negative charges along the ferroelectric axis. As it can be appreciated in the schematics of Figure 1a, the existence of spontaneous polarization gives rise to bound *polarization* charges at the crystal surfaces normal to the ferroelectric axis. The polarization charge density $\sigma_p$ verifies $\mathbf{P_s} \cdot \mathbf{n} = \sigma_p$, where $\mathbf{n}$ is an unit vector normal to the surface. This bounded polarization charge is usually compensated by outside *screening* charges as illustrated in Figure 1b keeping the neutrality at the crystal surface. The magnitude of the spontaneous polarization changes with temperature and disappears at a critical temperature called Curie temperature $T_C$. For LiNbO$_3$ (LN) and LiTaO$_3$ (LT), ferroelectrics of particular interest for biological applications, $T_C$ is very high (1170ºC and 680ºC respectively), far from operation temperatures in biological uses.

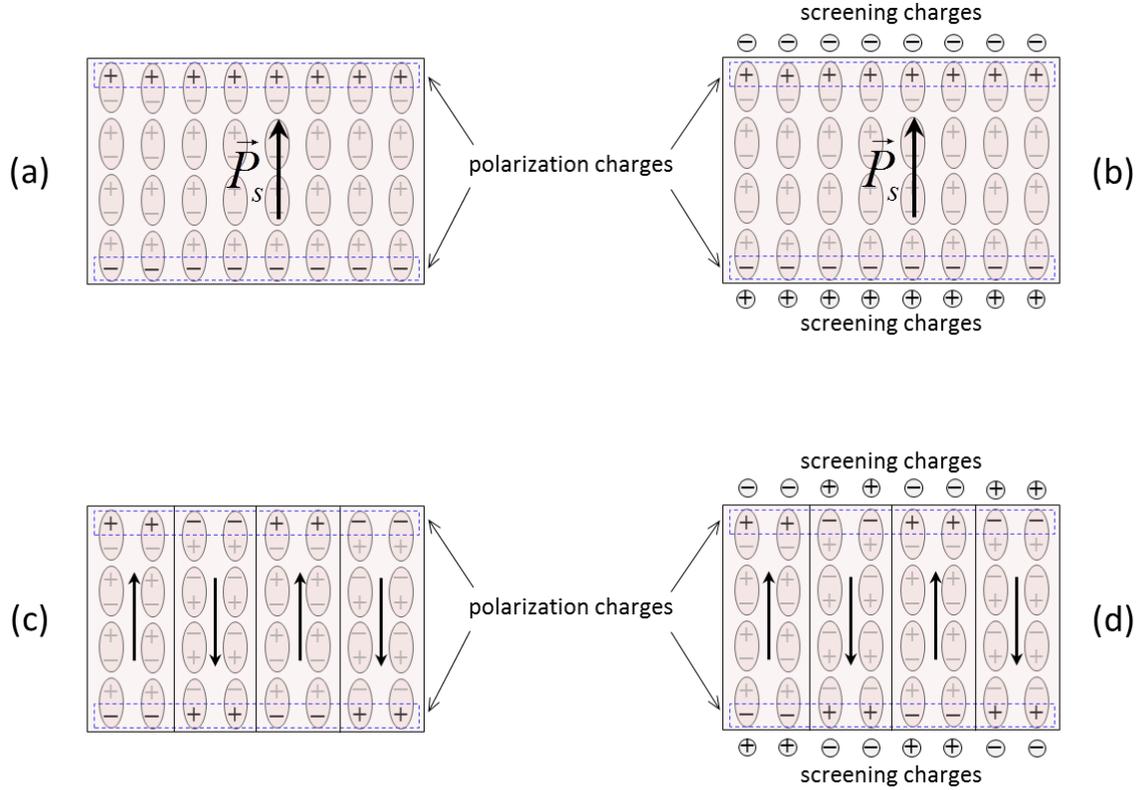

FIG. 1. Schematics of the cross section of a ferroelectric crystal, such as LiNbO$_3$ or LiTaO$_3$, showing the spontaneous polarization and the surface polarization and screening charges in several situations: (a) monodomain crystal with non-compensated polarization charge, (b) monodomain crystal with compensated polarization charges, (c) periodically poled crystal with non-compensated polarization charge and, (d) periodically poled crystal with compensated polarization charge.

The ferroelectric axis can be reversed in the whole material or in a region of it (domain) by applying an electric field whose strength is greater than a threshold value called coercive field. Therefore, a ferroelectric material can be monodomain (see Fig. 1(a), 1(b)) or polydomain (Fig. 1(c) and 1(d)). The boundaries between domains are usually called domain walls and, just there, an electric field appears.[17] By domain inversion, a variety of domain structures can be fabricated with many technological applications in different fields such as nonlinear optics, acusto-optics, nanoparticle trapping or biological applications.[12,18] Hence, domain inversion engineering has become a very active research field in the recent years. Particularly useful for all kind of applications are structures with periodically inverted domains (e.g. periodically poled LN -PPLN-), either 1D or 2D.

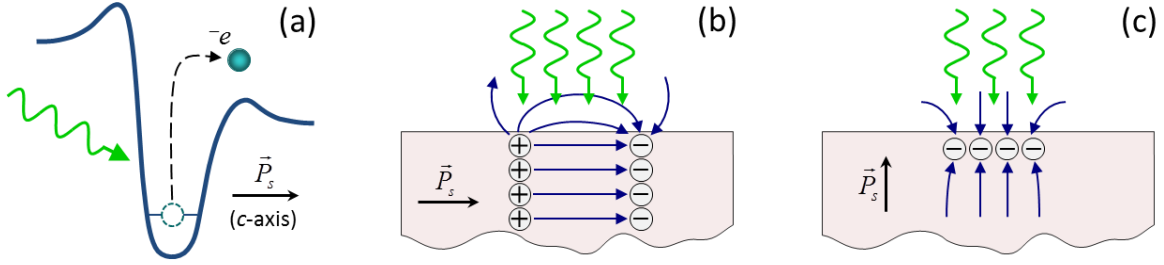

FIG. 2. Schematic diagram of: (a) the directional electron photo-excitation in PV ferroelectric crystals such as Fe-doped LiNbO₃, (b) the photo-excited charges and photovoltaic electric field when the active crystal surface is parallel to the polar axis (*x*-cut or *y*-cut), and, (c) the photo-excited charges and photovoltaic electric field when the active crystal surface is perpendicular to the polar axis (*z*-cut).

Ferroelectric materials present several effects able to generate internal electric fields. Among them, the bulk photovoltaic (PV) and the pyroelectric (PY) effects are particularly relevant for biological applications, the subject of this focused review. The PV effect appears in certain doped ferroelectrics, and it is singularly strong in Fe-doped LiNbO₃ (Fe-LN).[19] The effect arises from asymmetric photo-excitation of electrons from certain impurities (see Fig. 2(a)) due to the non-centrosymmetric crystal lattice, giving rise to an electric (photovoltaic) current along the polar *c*-axis. The PV current density induced by an illumination of intensity *I* can be written as:[12]

$$j_{pv} = e\alpha\Phi l_{pv} \tag{1}$$

where, $\Phi = I/h\nu$, is the photon flux, $\alpha$ the absorption coefficient and $l_{pv}$ the PV effective drift length ($\approx$1-5 Å). As a consequence, a light induced charge density appears and, correspondingly, a bulk electric field generates, which at steady-state conditions can be written as:

$$E_{pv} = j_{pv}/en\mu \tag{2}$$

*n* being the steady density of photo-excited electrons, $\mu$ the electronic mobility of the ferroelectric crystal and $j_{pv}$ the current density given by (1). It is worthwhile noting that when illuminating with a light pattern, spatially modulated electric fields closely correlated with the exciting pattern are obtained. For biological and technological applications of the PV effect two different geometrical configurations of the ferroelectric crystal has been used, as schematically represented in Figure 2(b) and 2(c): crystals with the active surface parallel (Fig. 2(b)) or perpendicular (Fig. 2(c)) to the polar axis, customarily called *x*- or *y*-cuts and *z*-cut, respectively.[20] In both cases, fringe electric fields appear outside the sample, as also indicated in these figures where the field lines and the PV-induced charges are drawn. Note that in Figures 2(b) and 2(c), polarization and screening charges, as illustrated in Figure 1, are not shown to simplify the schematics.

In turn, the PY response is just related to the surface polarization and screening charges existing in the ferroelectric surfaces. After a sudden temperature change $\Delta T$, caused either by heating or cooling, $P_s$ changes its magnitude, giving rise to an uncompensated surface charge density $\sigma_{py}$ at the $z$ crystal faces. $\sigma_{py}$ could arise either from polarization charges or screening charges, depending on the sign of $\Delta T$, and can be written as:

$$\sigma_{py} = \Delta P_s = c_p(T_s - T_0) = c_p \Delta T \qquad (3)$$

$c_p$ being the pyroelectric coefficient. As a consequence of this surface charge density $\sigma_{PY}$, a pyroelectric field develops in the surroundings of the ferroelectric crystal. For pyroelectric applications, $z$-cut crystals (see Fig. 2(c)) are mostly used.

Therefore, both the light-induced PV effect and the thermally induced PY effect generate surface charge densities and, correspondingly, induce PV or PY electric fields in the vicinity of the ferroelectric crystal. Additionally, as already mentioned, such electrical phenomena can be also found in the domain boundaries of poly-domain structures. It is worthwhile remarking that, in all these mechanisms, the electrical field is generated without either real electrodes or voltage suppliers, what is often referred as operation with "virtual electrodes". Moreover, in the case of the PV effect, the "virtual electrodes" can be design at will by using suitable light patterns for illumination. The surface charges and electric fields of ferroelectric crystals and structures are the basis of their outstanding applications as active substrates for trapping, patterning and other kind of manipulation effects on micro and nano-objects.[12,13,21-23] In the next sections, we will review applications reported in the biological world.

## III. BIOLOGICAL APPLICATIONS OF FERROELECTRICS

The field of biological applications of ferroelectrics is quite novel, practically going back just a decade if we do not include publications related to the piezoelectric effect. Classical uses for ferroelectrics, predating the biological applications, have fallen within the realms of optoelectronics, solid-state physics, optics and holography, and electronics to name a few. Here, we will be presenting the pioneer biologically-oriented works in chronological order. As mentioned in the Introduction, we will present the published results also attending to three main approaches to induce the electric field: domain structures, the PV effect, and the PY effect (see Section II). These mechanisms have been put to use to trap/arrange or modulate different biological systems. First, trapping/patterning will be introduced, followed by biological modulation experiments.

### A. Biological trapping and patterning

Ferroelectric trapping and patterning are the result of EP and/or DEP forces acting on the biological sample. Under the right conditions, electric fields can be induced in a ferroelectric material that extend into its surroundings as fringe fields.[12,24] This altered electric field, both in magnitude and direction, is the source for the EP and DEP phenomena. Biological structures (molecules, proteins, cells) respond to this EP and DEP forces, and can become trapped at certain regions on the ferroelectric surface. In what follows, biological trapping/patterning experiments will be presented attending to the different ferroelectric excitation mechanisms introduced above. Relevant experiments in relation to biological trapping appear summarized in Table I at the end of this section for easy consulting.

### 1. Domain structures

Ferroelectric domain engineering allows reversing the electric polarization to achieve very small domains (size < 1 µm).[24,25] Such polarization reversal gives rise to very localized and intense electric fields at the domain walls. The first report exploiting this domain structure effect to manipulate a biological sample was the work by Dunn *et al*.,[26] although there is a previous report on photoinduced trapping (see Section III.A.2 below). This group was able to trap and promote the assembly of tobacco mosaic virus particles over a $PbZr_{0.3}Ti_{0.7}O_3$ (PZT) thin film. The domain surface polarization was adjusted to the desired polarity by current flowing through an atomic force microscope tip. Counterintuitively at first, the viral particles collected at positive domains when they themselves display a positive net charge (see Fig. 3).[26] This was explained by the authors due to a double layer produced at the ferroelectric surface. At first, the positive ferroelectric domain attracted anions from the liquid which, on their part, attracted the (larger) viral particles and trapped them.

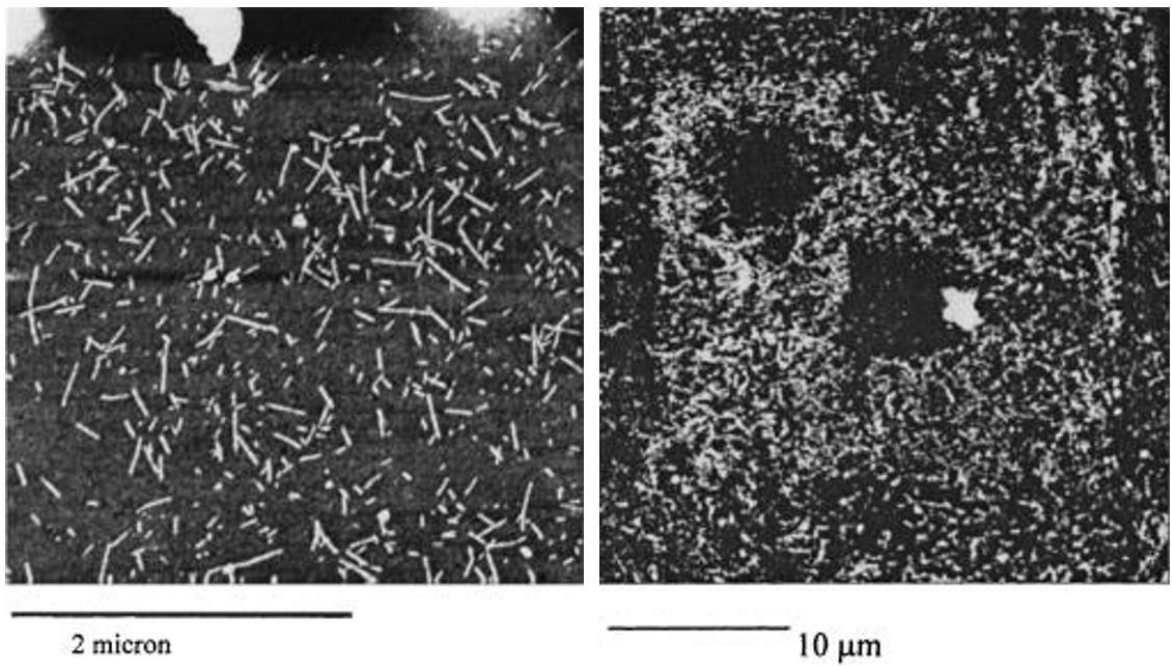

FIG. 3. Example of PZT ferroelectric trapping of tobacco mosaic virus proteins. On the left, normal distribution of virus particles on the surface of the unpoled ferroelectric. On the right, positively-poled regions trap and concentrate virus particles (more dense white dots distribution) meanwhile contiguous negatively-poled regions appear nearly vacant (black spots). Reproduced with permission from Dunn *et al*., Appl. Phys. Lett. **85**, 3537 (2004).[26] Copyright 2004 American Institute of Physics.

The next work on biological trapping making use of ferroelectrics appeared in 2012.[27] The ferroelectric employed was polyvinylidene fluoride (PVDF) as films. Corona discharge exposure led to electric poling of the β-isomer (β+ or β- surfaces vs. non-poled β or α isomers). It was shown that human fibronectin (FN), an important cell-substrate attachment protein, was significantly more adsorbed over the β+ or β- surfaces in comparison to the other surfaces. Additionally, work was reported on MC3T3-E1 cell cultures (mouse osteoblast precursor cell line) exposed for different times (3 and 7 days) to the PVDF surfaces previously incubated with FN. Somewhat surprisingly (see further reports below) the glass control surface showed the highest cell concentrations and densities in comparison to all PVDF surfaces. Some tentative explanation relying on differences between FN distributions and conformations due to electric charges was provided by the authors to explain the biological results.

Soon after Christophis *et al.* published more results on cell cultures.[28] The material employed was LT with periodically-poled domains obtained by submitting the ferroelectric to voltages of 10 kV. The domains had a period of ~ 22 μm. As biological models, rat embryonic fibroblasts, expressing a yellow fluorescent protein (REF52YFP), and human leukemia line (KG-1a) were seeded over the poled ferroelectric. The authors report that REF52YFP cells avoided the regions encompassing the domain boundaries, precisely where the field changes polarity and field lines are most concentrated.[24] However, the effect was subtle, and it was practically non-significant when the more motile KG-1a cells were employed. It is worth mentioning that fluorescent protein patterning over the ferroelectric was also assessed, but no pattern was observed. Then, a paper was published accounting on the role of ferroelectric face polarity on cell adhesion and morphology.[29] The material employed was z-cut LN and the cells NIH 3T3 mouse fibroblasts. Cell attachment was adequate (no differences to glass control surfaces) but the cell growth was somewhat faster over LN surfaces, in particular 24 h after cell seeding. Additional experiments carried out involved assessment of several cell modulation responses (morphology, wound healing test) with very interesting outcomes (see Section III.B.1 below). The same research group recently published two papers on this topic, in which the experimental model was the same but the cell adhesion was studied by total internal reflection microscopy holography.[30,31] The results show that cell contact areas are significantly smaller on the (+) face of the ferroelectric as compared to the (-) face or the glass control substrate (no differences between these last two). A process whereby differential (depending on the substrate polarization) adsorption of chemical species present in the cell medium takes place at the surface is advanced to explain the observed differences (see Sections IV.B and C below). The same argument has been put forward by other authors in relation to the use of certain inorganic materials (hydroxyapatite and electrically-active ceramics) to increase cell attachment, bonding and growth for bone implants.[32-34] In these publications a preponderant role is given to negative-charged surfaces to promote $Ca^{2+}$ cations adsorption, which favors extracellular matrix deposition and cell attachment. We will discuss this in more detail in Section IV.

More papers on this topic have been published in 2017. Kilinc *et al*. made use of PPLN to assess its suitability as a substrate for neuron attachment and directed axonal growth.[35] Primary embryonic mouse cortical neurons were seeded over LN (x-cut or z-cut), PPLN (stripe domains) or glass as a control substrate. Neurons displayed a more clustered arrangement and smaller axonal frequency over z-cut LN, no matter if (+) or (-) faces, in comparison to x-cut LN or glass. In stripe PPLN the axons showed a tendency to align with the domain boundaries, in contrast to the results of Christophis *et al*. where cells avoided domain boundaries.[28] Axons growing on etched LN (hexagonal pits) displayed a tendency to avoid these pits (see Fig. 4).[35] Hence, neuronal axons are able to sense changes in charge sign and topography of LN substrates and respond correspondingly. As the authors state, these substrates can be a promising platform for directed neuronal growth. Other cell lineages can potentially be steered into particular structures with a similar setup.

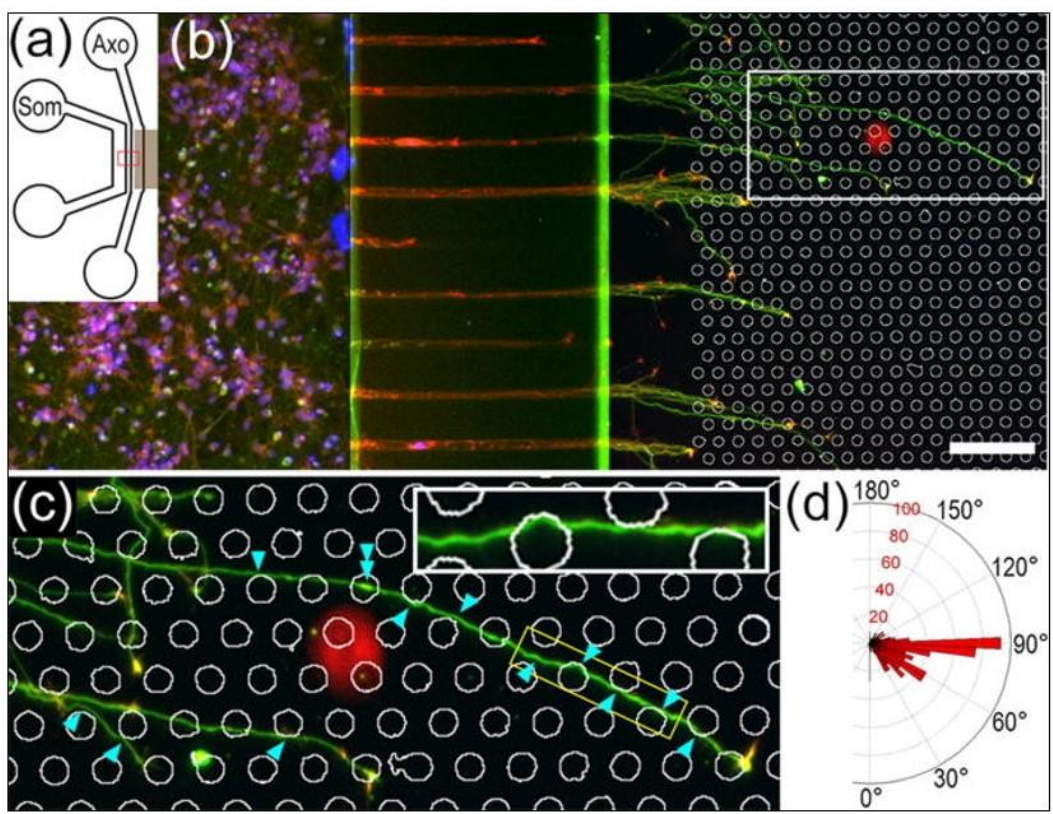

FIG. 4. Fluidic isolation of cortical axons on the etched LiNbO₃ substrate. (a) Layout of the bicompartmental microfluidic chip consisting of the somatic (Som) and axonal (Axo) chambers. The brown area indicates the location of the aligned etch pattern. (b) Microfluidic neuron culture on the etched y-cut LiNbO₃ substrate (red boxed area in (a)), showing β3-tubulin (green), F-actin (red), and nuclei (blue), superposed with the etch pattern (white). Scale bar = 100 μm. (c) Marked area in (b) is 2.6x magnified to show the deformations in axon shafts. Axons follow the edges of microstructures (arrowheads). Occasionally, when an axon shaft crosses through a pit, increased cytoskeletal density was observed (double arrowhead). Inset: the axon segment in the yellow box is 2x magnified. (d) Angular histogram of axon segments on the entire patterned region in the axonal

chamber (numbers of axon segments given in red). Axon segments are typically aligned at 0° or 30° with respect to the microchannels (90° in the histogram) as they navigate through the hexagonal pits pattern. Reproduced with permission from Kilinc *et al*., Appl. Phys. Lett. **110**, 053702 (2017).[35] Copyright 2017 American Institute of Physics.

A very interesting approach is that reported by Toss *et al*. in which human primary fibroblasts were seeded on a ferroelectric polymer, polyvinylidene fluoride trifluoroethylene (PVDF-TrFE).[36] Under adequate external electric fields this polymer can be poled. The fibroblasts grew perfectly well on PVDF-TrFE for several days. After they have formed confluent 2D films, an external electric field was applied, with the consequence that large fibroblasts "clusters" or "film islands" detached from the substrate because of polymer surface sign reversion. The authors envision this as an efficient method to provide cultured tissues for transplants or skin grafts, without engaging in mechanical or chemical cell detachment, which are more challenging for the cells and frequently destroys any 2D or 3D tissue structuration obtained during the *in vitro* culture.

## 2. Photovoltaic effect

Here results published in relation to biological trapping through the use of the PV effect, also known as photovoltaic optoelectronic tweezers,[12] will be presented. The first proposal to take advantage of an increase in photoconductivity due to spatially-modulated light patterns (moving gratings) to trap/move matter in liquids or gas was made by Kukhtarev *et al*. in 1998.[37] The paper is theoretical, but it can be read in the introduction: "We suggest the use of a novel optical technique based on light-induced gratings in hazardous materials existing in aerosols in liquid or vapor phases. This method can be used for detection, separation, and removal of undesirable materials and/or microorganisms." Therefore, the possibility to put to use photoinduced gratings in certain photonic materials to trap physical (biological) particles was advanced. Shortly after, this group reported the first experimental trapping effect employing ferroelectrics (z-cut Fe-LN) on *Escherichia coli* cells.[38] Just an image of poor quality is provided, but nevertheless it represents the first attempt to trap biological objects by making use of photoinduced gratings (532 nm laser light). This work also provided evidence that living fungal cells (*Aspergillus flavus* and *Aspergillus niger*) can grow on ferroelectrics highlighting their biocompatibility. Additionally, this paper reports some biological modulation (see Section III.B.2 below).

More than a decade elapsed before new experimental results on the trapping of biological matter due to optical ferroelectric excitation was published in 2015 by our group.[39] Both x-cut and z-cut Fe-LN were employed to arrange 1D (x-cut) and 2D (z-cut) patterns of different biological materials: fungi spores (~ 10 µm), and pollen grains (~ 70 µm) and fragments thereof (~ 1-10 µm) in hexane or air. The photoinduced gratings in the LN were excited by interfering patterns of 532 nm laser light. These results proved that it is possible to arrange microscopic biological materials with a very high spatial accuracy (~ microns). More experiments in this line were done and the results were published the next year.[40] Some of the obtained patterns are shown in Figure 5. The mechanism acting on the bio-particles and trapping them is DEP, as they are electrically neutral in practice. In fact, the induction of DEP as a result of the photovoltaic effect is a phenomenon already described with inorganic materials and particles.[20,41,42] The electric field can reach very high values ($\sim 100 \, \text{kV cm}^{-1}$) and it is quite inhomogeneous close to the photoexcited ferroelectric surface.[43,44] These are the ideal conditions for DEP transport which relies on field inhomogeneity to act. Charged particles, if present, will also react to the field, and even accumulate in different regions depending on their own charge and the substrate polarity/charge pattern .[45] So it is important to remark that ferroelectric trapping can drive both EP and DEP,[46] the final patterning obtained depending on the electric charge of the entity actuated upon.

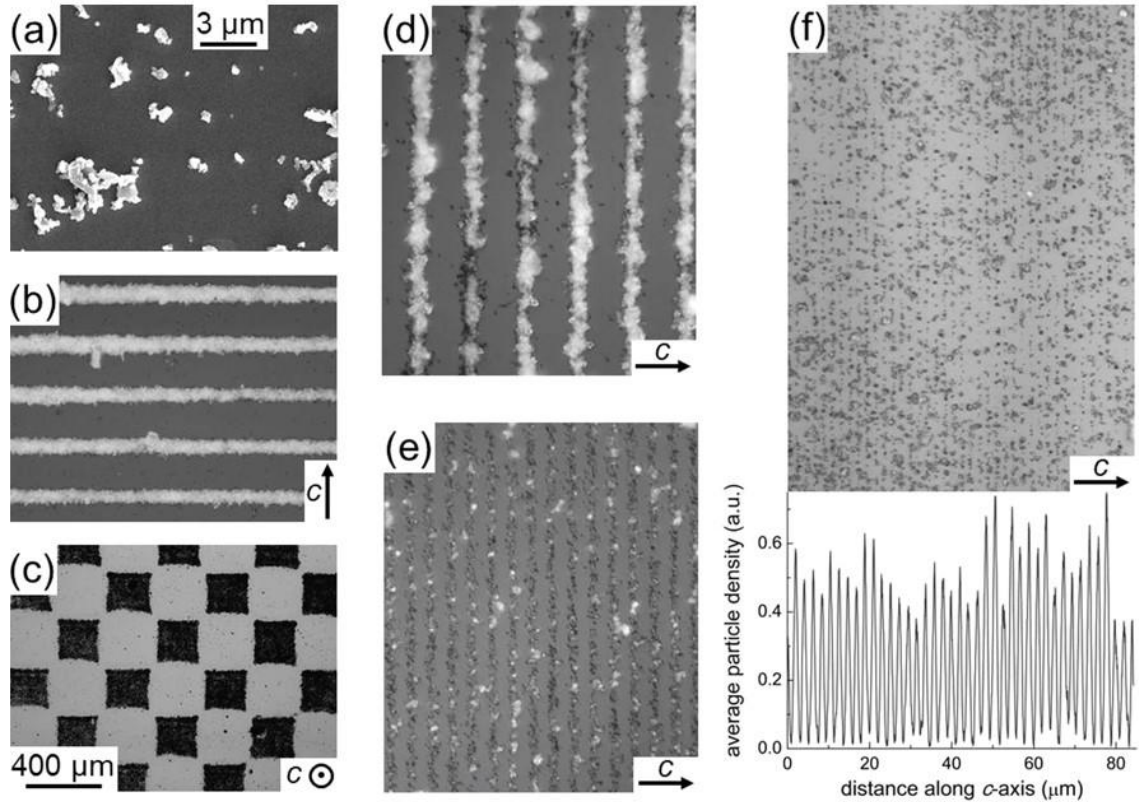

FIG. 5. (a) SEM image of Himalayan cedar pollen fragments. (b) Microscope image of a 65 µm period pattern of these pollen fragments obtained with sinusoidal illumination of the same spatial period. (c) Microscope image of a 2D pattern of the same pollen fragments obtained after illumination with a mosaic of squares with 200 µm side. Microscope images of periodic patterns with decreasing periods: (d) $\Lambda = 20$ µm, (e) $\Lambda = 8$ µm, and (f) $\Lambda = 2$ µm. In the bottom of the last one (f), the corresponding average particle density profile along the c-axis direction is also shown. Reproduced with permission from Jubera *et al.*, Appl. Phys. Lett. **108**, 023703 (2016).[40] Copyright 2016 American Institute of Physics.

An important leap forward was reported by Miccio *et al*. who were able to trap living bacteria (*E. coli*) in a water-based cell medium.[47] This is very important because the water molecule shows a permanent electric dipole, thus water screens effectively any charge in submerged substrates. Also, biologically sustaining media must include a series of salts and charged molecules (aminoacids, proteins) to preserve cell isotonicity and provide the necessary metabolites for cell metabolism. In consequence, ferroelectric trapping of living organisms presents a particular challenge (charge and field screening), which is currently being dealt with by several research groups. In the mentioned experiment, the bacteria were trapped on a x-cut Fe-LN excited with an Ar ion laser (514 nm) to produce interfering patterns (12.5, 25, 50 and 100 µm) within the crystal. In all cases bacteria displayed a marked reduction of their Brownian motion and were largely aligned close to the orthogonal (60° or more) in respect to the photoinduced fringes. This means the bacteria were aligning themselves with the electric field gradient displaying a "dipole-like" behavior. In connection with this, a torque phenomenon has been reported very recently upon cylindrical zeolite microparticles, trapped on a similar Fe-LN substrate and subjected also to interfering illumination (532 nm).[48] The zeolites (1.5 µm in diameter x 4 µm in length) arranged following the electric field lines, simultaneously experiencing DEP and torque to do so. The bacteria in the experiments mentioned have very similar dimensions: 0.5 µm in diameter x 2 µm in length. Even more, bacteria incubated for 2 h after trapping were capable of cell growing while aligned. As a result, aligned cylindrical bacterial colonies were obtained, as featured in Figure 6. So, these experiments confirmed that 1) cell viability was preserved for at least 2 h after trapping, and 2) that cells were experiencing mild enough conditions so as to proliferate while trapped. This is an important result, to be put in connection with the experiments discussed later in Section III.B.

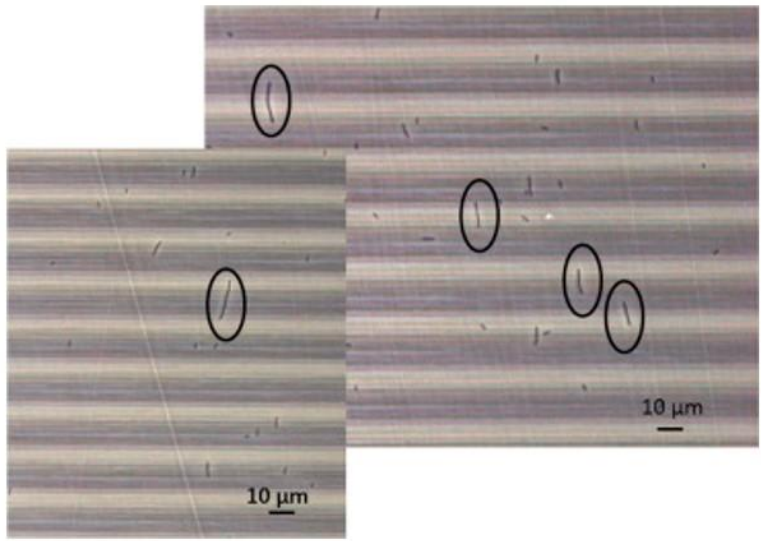

FIG. 6. Two pictures of bacterial (*E. coli*) chains longer than 10 µm aligned perpendicularly to the grating planes over Fe-LN. Photoinduced gratings ($\Lambda = 25$ µm) recorded before bacteria exposure to the material. Notice the important chain elongation considering the size of a single bacteria (~1 µm). Reproduced with permission from Miccio *et al*., Opt. Lasers Eng. **76**, 34 (2016).[47] Copyright 2015 Elsevier.

An interesting novel biotechnological application has been advanced by Elvira *et al*., in which the PV effect is employed to generate a pattern of silver nanoparticles on a Fe-LN substrate.[49] Then, these Ag nanoparticles serve as plasmonic platforms upon which the fluorescent emission of several molecules is enhanced. Among the biomolecules tested, both fluorescein-labelled DNA and synthetic peptide nucleic acid showed an increased fluorescence signal on the patterned Ag nanoparticles. This opens possibilities for biomolecule detection and analysis at very low concentrations due to this plasmonic enhancement.

### 3. Pyroelectric effect

In contrast with the two previous approaches to use ferroelectricity for biological trapping/arranging, pyroelectricity has been very scarcely applied directly for such a purpose. However, there are a few precedents in microparticle trapping, and at least one report of cell trapping, which makes the approach appealing. In 2008 results were published showing that flour microparticles (1-10 µm) in oil could be trapped on a PPLN (hexagonal domains) after heating to 100 ºC for 1 minute.[21] The particles clustered at the domain boundaries, as would be expected for a DEP trapping mechanism.[50] Further confirmation of the DEP was obtained employing latex particles (1 µm) in carboxyl acid. A few years later the same group reported trapping of yeasts cells on a very similar PPLN substrate.[22] However, the biological sample was dispersed in paraffin oil and the temperatures are assumed to be out of the biological tolerance range. Nevertheless, the results prove again that biological structures (not necessarily alive) can be efficiently trapped by ferroelectric materials adequately excited.

This group has recently published an interesting use of the pyroelectric effect to pole a polymer film on which living cells are seeded at a later stage.[51,52] The polymer film is cast over the ferroelectric substrate, PPLN in this case (stripes or hexagonal domains). Controlling the temperature of the system it is possible to induce the pyroelectric effect in the substrate and, at the same time, reach a temperature above the glass transition temperature of the polymer. Then the molecules in the polymer become poled by the pyroelectric field beneath the film. On cooling, the polymer structure becomes rigid again, replicating the pyroelectric pattern. The film is detached from the substrate and cells are seeded on it. The microscopic images clearly show how the cells avoid positively charged film regions while attach to negative spaces. Very recently this group has published on the trapping of living bacteria (*E. coli* and *S. epidermis*) which leads to an efficient biofilm growth, with a very high bacterial viability, on polymer surfaces electrically poled by previous exposure to the pyroelectric effect of LN.[53] The polymers (polysulfone and polystyrene) were prepared as strips or fibers and poled by exposure to the pyroelectric action of LN. Bacterial cultures subsequently exposed to the poled polymers favored the growth of biofilms on them as compared to controls. The authors argue that the efficiency and biocompatibility of the approach is due to a less-interfering action of charge below the polymer surface, which does not disrupt the ionic composition of the bacterial outer membrane.

Another step in this direction by this group is the publication of a setup to arrange cell growing spots "at convenience", where biologically-friendly polymers are "printed" on a substrate making use of the pyroelectric effect of nearby LT.[54] In this way, it is possible to grow cell colonies in particular spots or patches, in different geometries or even grow individual cells in tiny "islands". This is a further bio-oriented advancement in the so-called "pyroelectrodynamic shooting" technique developed by this research group some years ago.[55]

Finally, an aspect deserves remarking in relation to the interplay between the photovoltaic and pyroelectric effects. When a ferroelectric is optically excited (*e.g.* Fe-LN) some fraction of the light energy is degraded to heat during the absorption. As the pyroelectric effect is inherent in ferroelectrics, this photothermal side effect activates pyroelectricity in the illuminated sample. Once acknowledged, this synergy between photovoltaic and pyroelectric aspects can be put to use to achieve more defined or stronger "virtual electrodes" for different applications.[56] In this line, in 2008 a group successfully trapped polystyrene nanoparticles (100 nm) on PPLN (stripe domains) making use of UV (254 m) ferroelectric excitation modulated by the pyroelectric effect.[57] The important line here is that the experiments were done in water and the polystyrene nanoparticles were negatively charged (as can be expected from living cells) due to surface carboxyl-group functionalization. Thus, the trapping mechanism was EP in this case, but showed that there is an interplay between the photovoltaic effect and the temperature.

TABLE I. Summary of biological trapping and patterning using ferroelectric materials.

| Biological material | Ferroelectric material | Excitation mechanism | Remarks | Reference |
|---|---|---|---|---|
| *E. coli* bacteria | Fe-doped $LiNbO_3$ | z-cut photoexcited with 532 nm laser light | E.coli cells redistributed under laser interference gratings. | 38 |
| Tobacco mosaic virus particles | $PbZr_{0.3}Ti_{0.7}O_3$ | Domain structure | Viral particles concentrated to the point of self –assembly. | 26 |
| Fibronectin and MC3T3-E1 (mouse osteoblast precursor) | PVDF | Corona domain poling | Increased fibronectin adsorption | 27 |
| REF52YFP (rat fibroblasts) KG-1a (human leukemia cells) | $LiTaO_3$ | Periodically poled domains | Cells avoided domain boundaries (subtle effect). | 28 |
| Yeast cells | Periodically poled $LiNbO_3$ | Pyroelectric effect by hotplate (z-cut with hexagonal | Cells probably dead due to apolar solvent (paraffin oil) | 22 |

| | | domains) | and high temperature (100 °C) | |
|---|---|---|---|---|
| NIH 3T3 mouse fibroblasts | LiNbO$_3$ | z-cut mono-domain crystal | Adequate cell attachment to ferroelectric substrate. | 29 |
| Fungi spores and pollen grains | Fe-doped LiNbO$_3$ | Photovoltaic effect (532 nm laser light) | 1D (x-cut) and 2D (z-cut) patterning was achieved down to micron scale. | 39,40 |
| *E. coli* bacteria | Fe-doped LiNbO$_3$ | Photovoltaic effect (514 nm laser light) | 1D (x-cut) bacterial patterning and growing in water-based medium. | 47 |
| NIH 3T3 mouse fibroblasts | LiNbO$_3$ | z-cut mono-domain crystal | Cell focal contact areas larger on substrate (-) face than (+) face. | 30, 31 |
| Primary embryonic mouse cortical neurons | LiNbO$_3$ | Periodically poled domains (stripes and hexagons) | Neurons more clustered over z-cut LN than over glass. Axons follow domain boundaries on etched LN. | 35 |
| Primary human fibroblasts | PVDF-TrFE | Externally poled | Extensive 2D cell films gently detached from substrate | 36 |
| *E. coli* and *S. epidermis* bacteria | LiNbO$_3$ | Pyroelectric effect poling of polymers which then interact with bacteria | High viability bacterial biofilm establishment and growth | 53 |

## B. Cell response modulation

Several papers have dealt with modulation of biological systems through ferroelectric interaction, although they are less frequent than trapping related papers. They will be introduced in chronological order in what follows. As in the previous section, the results are presented attending to the method of ferroelectric excitation. Additionally, relevant biological modulation experiments are summarized in Table II at the end of this section for the interested reader.

### 1. Domain structures

Polarized domains also exert a modulation effect on cells in addition to a trapping action. Reports on cell modulation through polarized domains started in 2015. Human adipose stem cells cultured over FN-covered corona-poled PVDF displayed more focal adhesions and cell substrate attachment area when they were interacting with β- surfaces.[58] Cell size was smaller when interacting with PVDF (irrespective of the poling) in comparison to control cells growing on glass. In addition, stem cells differentiated into an osteogenic phenotype in a similar way on polystyrene (positive control substrate for differentiation) as in β+ or β- PVDF. Craig Carville *et al*. describe that MC3T3 osteoblast cells grow faster either on (+) or (-) z-cut LN compared to x-cut LN or glass as a substrate control.[59] As can be seen in Figure 7 (a) cells covered the imaged surface faster on (–) z LN or (+) z LN during the first 3 days (trend continues up to day 11). They also measured the mineralization degree of the cells (osteoblasts are bone-synthesizing cells) through the alizarin red test. They found that, indeed, this cellular function was enhanced in cells growing on the (+) or (-) surfaces of LN. The authors hypothesize that preferential ion accumulation at the charged LN surfaces somehow stimulated the cell growth and metabolic activity. This ionic crowding will be discussed as one of the tentative biological action mechanisms in Section IV.C.

Practically at the same time, another group published very similar results.[60] The ferroelectric explored was again LN (z-cut vs. x-cut) and the cell model bone marrow stem cells from rats. They found that there was a slight increase in proliferation in z-cut (+ or -) LN surfaces compared to x-cut, but it was not statistically significant. However, cells were more spread (larger cell surface area) over (+) z LN in comparison to (-) z or x-cut (this is in conflict with most reports, where cells are more spread on the (-) z surface). An example is shown in Figure 7 (b), were cells on (+) z are more spread than in the other two images. Several osteogenic biomarkers were assessed, and it was found that the (+) z face was clearly stimulating osteogenesis by upregulating protein levels and genetic transcription. These elements indicate, in the opinion of the authors, activation of the TGF-β and Wnt signaling pathways, involved in osteogenesis in osteoblasts. Hence, (+) z LN was favoring differentiation of bone marrow stem cells towards an osteoblastic phenotype. Authors argued that electrostatic ionic arrangement (Stern layer) at the LN surface was the key element that actuated the observed biological response.

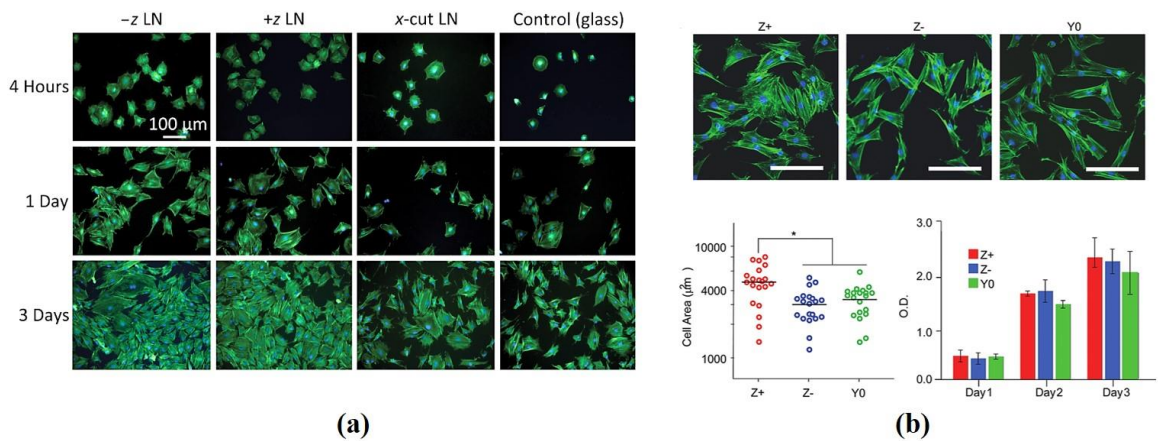

**(a)**                                                                    **(b)**

FIG. 7. (a) Representative fluorescent images of MC3T3 cells (nuclei and actin are stained with DAPI and phalloidin, respectively) on the first 3 days of culturing on −z, +z LN, with x-cut LN and glass cover slip as control samples. Cultures are more confluent on z-cut (+ or -) LN in comparison with the controls. Reproduced with permission from Craig Carville *et al.*, J. Biomed. Mater. Res. Part A **103A**, 2540 (2015).[59] Copyright 2014 Wiley Periodicals. (b) Spreading and proliferation of rat bone marrow stem cells on differently charged LN surfaces. (Top row) Actin was stained to visualize the cell spreading. F-actin was stained by Alexa fluor 488-conjugated phalloidin (green) and cell nuclei were stained by DAPI (blue). Scale bar: 200 μm. (Bottom left) Measurement of cell area for rat bone marrow stem cells in response to different surface charges after 24 h of culture. Data represented the mean ± standard deviation ($n = 20$, * $p < 0.05$). (Bottom right) WST-8 viability assays were performed to detect cell proliferation in the first 3 days ($n = 3$). Reproduced with permission from Li *et al.*, Adv. Healthcare Mater. **4**, 998 (2015).[60] Copyright 2015 Wiley-VCH Verlag.

Work by Marchesano *et al.* (already discussed in regards to trapping in Section III.A.1 above) showed that fibroblast growth was slightly enhanced over (+) or (-) z-cut LN compared to glass, in particular 24 h after cell seeding.[29] Cells were more spread on (-) z-cut LN, and displayed better defined actin stress fibers and vinculin expression than on the (+) z surface (see Fig. 8 (a)). If challenged by the wound healing test, fibroblasts closed the scratch faster on LN as compared to glass (see Fig. 8 (b)), with a slight advantage on the (+) face. All in all, it seems that the fibroblasts attach and spread better on the (-) surface, but recruit faster to heal the cell film at the (+) face, perhaps due to a decreased substrate attachment which permits a quicker cell migration. As it is the general trend, the authors theorize that charge-driven selective ionic adsorption on LN, in particular $Ca^{2+}$ "trapping" at the (-) face, explains the different cell behaviors observed for the (+) and (-) surfaces.

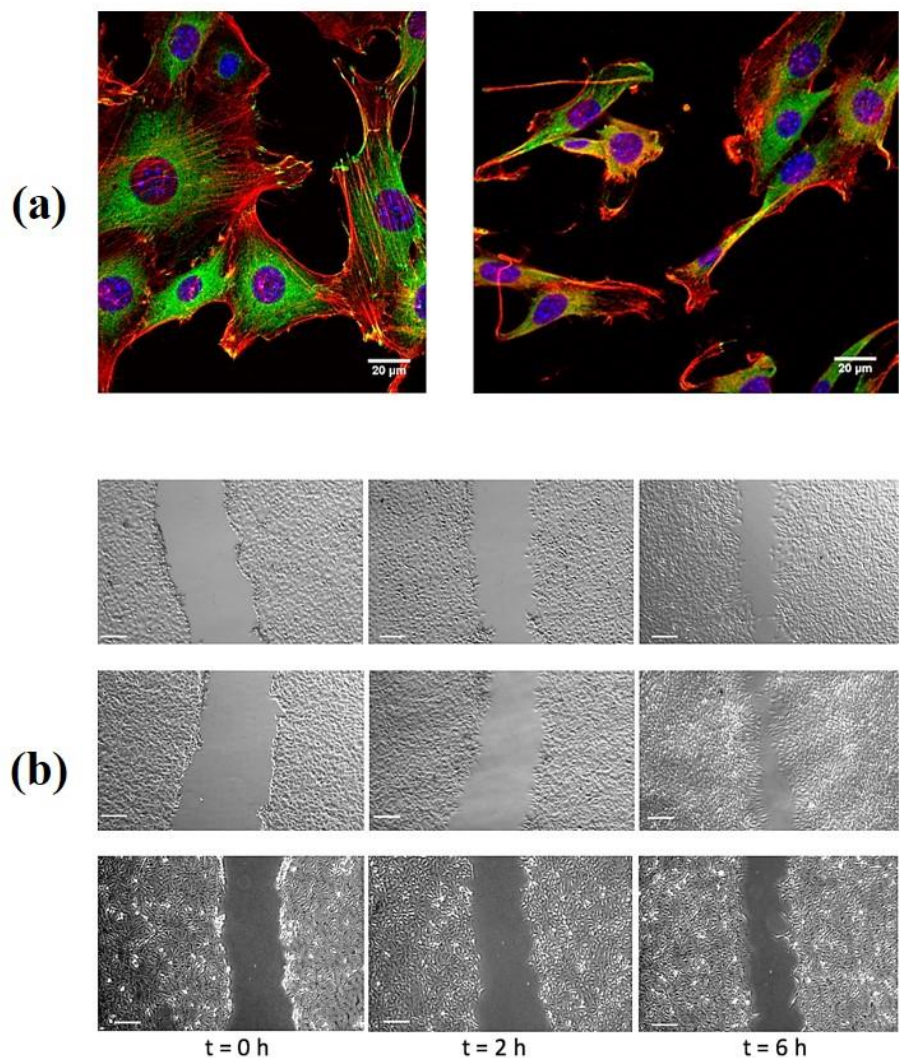

FIG. 8. (a) Confocal microscope images of NIH 3T3 fibroblasts plated on LN (left) on the c− face and (right) on the c+ face, after incubation for 24 h; nuclei, actin, and vinculin stained with DAPI, TRITC-conjugated phalloidin, and DyLight488 vinculin, respectively. Notice the more spread cells on the c- face, with a more robust actin cytoskeleton and highly ordered stress fibers. (b) Phase contrast microscope images of the evolution of the scratch (wound healing assay) at 0, 2 and 6 h for NIH 3T3 fibroblasts plated on LN c− (top row), LN c+ (middle row), and glass slide (bottom row) as a control. Both cultures on LN heal the wound faster than on glass; but the fibroblasts on the c+ show the highest migration (scale bar = 100 μm). Reproduced with permission from Marchesano *et al*., ACS Appl. Mater. Interfaces **7**, 18113 (2015).[29] Copyright 2015 American Chemical Society.

Similar results have been reported on Saos-2 osteoblast cells, where cell spreading and cell proliferation were found higher on z-cut LN (+ or -) compared to a glass substrate.[61] Additionally, β1-integrin and vinculin expression, as well as alkaline phosphatase activity, were increased on LN vs. glass. The (+) z surface effect is accounted for due to selective (negatively charged) protein adsorption, while the (-) z surface action can be due to selective $Ca^{2+}$ attachment. Zhou *et al.* made use of corona discharge-poled PVDF over a titanium substrate to study the osteogenic differentiation of bone marrow stem cells.[62] They found that negatively-poled PVDF (β- PVDF) increased both cell proliferation and osteogenic phenotype (alkaline phosphatase, collagen I and osteopontin expression) as compared to unpoled PVDF-Titanium.

A different approach has been taken by Li *et al.* where the ferroelectric is employed as nanoparticles (NPs) that functionalize an organic molecular scaffold structure.[63] Fibers of poly-(L-lactic acid), either forming random coils or aligned, incorporated in some cases NPs of $BaTiO_3$ (~100 nm) in different proportions. Apart from studying several physical and structural parameters, the fibers were assessed for biological activity on rat bone marrow stem cells. It was found that there was an enhancement in cell proliferation, particularly at day 3, when cells were growing on fibers (random or aligned) with $BaTiO_3$ NPs. Also, on random fibers plus NPs there were statistically significant increments of RUNX-2 (a transcription factor for osteogenesis) levels and alkaline phosphatase activity in comparison to the other conditions. Tentatively, the authors ascribe a mechanism for the observed biological action on the activation of the RhoA/ROCK signaling pathway, part of the Rho GTPases family of signaling proteins, due to electrotaxis-derived changes to the cytoskeleton and cell morphology (see Section IV.B).

A step forward has been taken by Liu *et al.* by studying a ferroelectric substrate ($BiFeO_3$ –BFO-) in an *in vivo* bone regeneration model.[64] Rats with a bone lesion in the femur received implants of neutral $SrTiO_3$ (control) or $SrTiO_3$ functionalized with BFO(+) or BFO(-) nanofilms. Bone regeneration in terms of better implant contact and regenerated volume were statistically significant for both BFO-covered implants (best results observed with BFO(+)). Further, *in vitro* experiments showed enhanced cell attachment, spreading and osteogenic differentiation of mesenchymal stem cells over BFO nanofilms.

Tang *et al.* have just reported a detailed work on the cell adhesion enhancement effect of PVDF-TrFE upon MC3T3-E1 cell cultures.[65] The polymer was poled by direct electric contact (as opposed to corona poling described previously) in a Cu-Ti capacitor arrangement, which provided positively charged surfaces for the biological experiments. Cells were seeded on the substrate, and assessment took place at several time points afterwards (up to 21 days in some cases). These included cell attachment and proliferation, actin cytoskeleton study, osteogenic differentiation (expression of alkaline phosphatase, collagen I and osteocalcin, plus alizarin red staining test), gene activation patterns (RUNX-2 and OCN, among others), and protein expression levels of integrin, phopho-FAK (focal adhesion kinase) and phopho-ERK (extracellular signal-regulated kinase). As a general trend, the more polarized the PVDF-TrFE substrate, the more marked the biological response towards enhanced proliferation and osteogenic differentiation at longer times. However, the most charged film (surface potential 915 mV) was less efficient than a moderately charged one (391 mV). Fibronectin was more abundant and interacted better with cells, through membrane integrins, on moderately charged films.

Recently the cytotoxicity of several ferroelectric ceramics, titanates of diverse composition, has been studied in primary mouse fibroblasts.[66] In some lead-free titanates cell proliferation was significantly above the control levels, pointing to a stimulating environment for the cells. The authors claim that ferroelectric wettability is an important factor to account for the biological activity observed, in relation to protein adsorption on surfaces. The more hydrophobic the substrate, the less biologically active in these experiments. Although the ceramics used were unpoled, the authors argue that at the very small limit spontaneous electric domains exist that determine the overall wettability and, hence, the biological action.

### 2. Photovoltaic effect

The photovoltaic/optoelectronic tweezers concept has also been applied to cell modulation. In fact, the first genuine reports on the biological action of ferroelectrics have been published under this experimental approach. The already cited paper by Kukhtarev *et al.* in 2002 also noted that fungal growth was enhanced on photoexcited (532 nm) z-cut Fe-LN, although the quality of the images is not very good and the statements are qualitative (no comparisons were made among different conditions).[38] Nevertheless, these authors were the first to suggest the use of ferroelectric materials for biological modulation.

Then, in 2011, our group published the first paper in which quantitative data related to the biological activity of ferroelectrics was provided.[67] We found evidence of the cytotoxic action of photoexcited x-cut Fe-LN. HeLa cells growing on the crystal and exposed to blue (436 nm) or green (546 nm) incoherent light (high pressure Hg lamp) rapidly developed structural alterations fully compatible with osmotic shock.[68,69] Indeed, total cell death was obtained with blue light (16 mW cm$^{-2}$ and dose 28.8 J cm$^{-2}$), in other words very mild conditions compared to typical laser irradiances (kW cm$^{-2}$ or higher). Green light was less efficient than blue light to induce cytotoxicity, in agreement with the absorption spectrum of Fe-LN.[67] Cell death was necrotic, with large plasmatic membrane bubbles (see Fig. 9). Initially we presumed that water electrolysis could be having a role in the observed cell death. Thus, we measured reactive oxygen species (ROS) generation under Fe-LN illumination ($H_2O_2$ is an abundant side product of $H_2O$ electrolysis). However, we were unable to find any $H_2O_2$ down to our detection threshold of ~1-5 µM. This was too low for the observed fast and massive necrotic death displayed by the cell cultures. Therefore, we concluded that ROS generation was not the main cell death mechanism, although very high local ROS concentrations cannot be discounted on basis to our results (see Section IV.A below for additional comments). Our experiments did not provide any direct clue as to the cytotoxic mechanism at work, but we are confident in that the mechanism inducing the cell damage and death is, most probably, membrane electroporation (see Section IV.D). Remarkably, Fe-LN microparticles (~1-3 µm), obtained by grounding a wafer piece, were also cytotoxic under similar light exposure, although longer illumination times (60 min) were necessary to induce necrosis. Further experiments revealed that Fe-LN NPs (~ 100 nm) displayed a certain cytotoxicity under green light.[39] Unfortunately, the NPs by themselves (no light exposure) showed also a high cytotoxicity.

Ongoing research by our group is currently trying to assess the cell damaging mechanisms in depth.

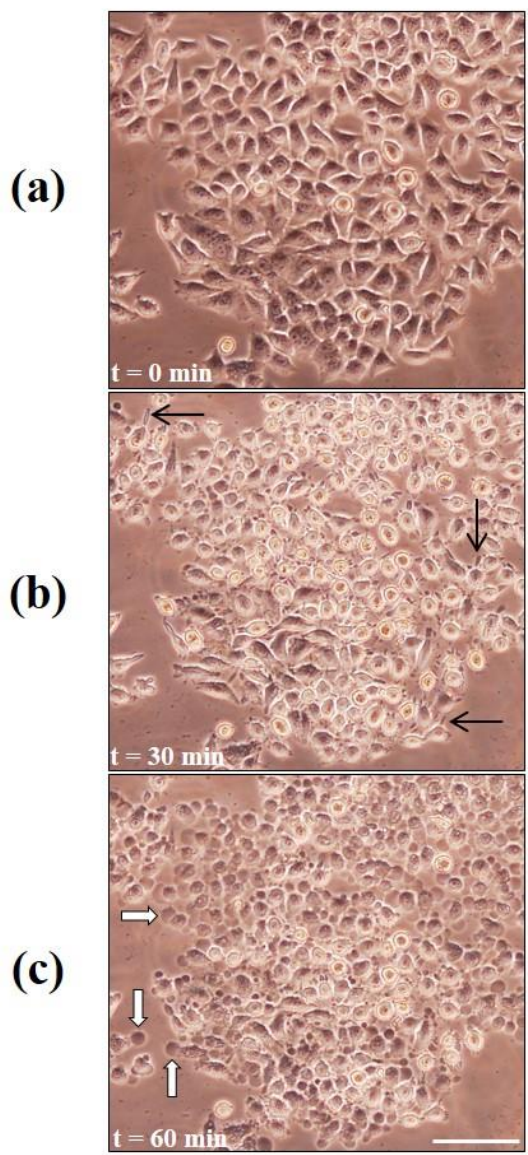

FIG. 9. Necrotic cell death induced on HeLa cells growing on a Fe-LN crystal under 546 nm light exposure (59 mW cm$^{-2}$). (a) Cell culture before light exposure displaying a normal cell morphology. (b) Same cells at the end of a 30 min light treatment. Notice the drastic morphology change to a round shape and the remarkable increase in cell refringence. Membrane structures (black arrows) still appear attached to the crystal. (c) Cell necrosis 60 min after starting light exposure. Large membranous bubbles (white arrows) can be seen budding from each cell. Many cells display more than one protruding bubble. The nuclear area appears blackened and fibrous-like. Scale bar = 100 μm.

It is important to remark at this point that several recent papers claim that no assessment on the biocompatibility of ferroelectrics has been done previous to 2012. This is incorrect, as publications by Kukhtarev *et al*. [38] and Blázquez-Castro *et al*. [67] (including an active patent filled in 2009 and granted in 2012)[70] already have provided experimental proof that living cells are perfectly capable of growing and proliferate on Fe-LN for extended periods of time -several days in our experiments-.

### 3. Pyroelectric effect

Finally, to the best of our knowledge, only two papers have been published making use of the pyroelectric effect for biological modulation. The first one, published in 2012, reported a remarkable bactericidal action of LN and LT microparticles (<5-15 µm) in an *E. coli* model.[71] Excitation of the pyroelectric effect was achieved by thermally cycling between 20 and 45 ºC for 6 h. Both LN and LT displayed bactericidal activity, the smaller the microparticle the more effective the result obtained. In contrast to our results,[67] the authors found a measurable oxidative activity under the experimental conditions, as assessed by the increased fluorescence of the dichloro-dihydro-fluorescein-diacetate ROS dye probe. Thus, they theorized that the cytotoxic mechanism had to do with an increase of ROS presence in the cell medium, presumably due to pyroelectrically-driven microparticle-catalyzed splitting of water.

There is a single, very recent report on eukaryotic cell modulation using the pyroelectric effect.[72] In this work, tourmaline (a natural pyroelectric mineral) microparticles (3 µm) were subjected to different temperatures, both above and well below room temperature, before mixing them with a cell suspension (Chinese hamster lung cell line DC-3F) at room temperature. Afterwards, the cell suspension was exposed to bleomycin, a powerful cytotoxic compound that does not penetrate the plasmatic membrane unless it is somehow compromised. The results showed that cell death due to bleomycin was observed when cells were exposed to tourmaline, and that cytotoxicity was directly related to the temperature change the tourmaline underwent when it was mixed with the cell culture. In summary, some degree of cell membrane poration was seemingly taking place in these experiments, and the evidence points to the pyroelectric effect of tourmaline as the causal agent. This is to be connected to our 2011 results, where membrane poration was presumed to be the most probable necrosis induction mechanism (see Section IV.D).[67]

TABLE II. Summary of biological modulation using ferroelectric materials.

| Biological material | Ferroelectric material | Excitation mechanism | Remarks | Reference |
|---|---|---|---|---|
| Fungal cells | Fe-doped LiNbO$_3$ | z-cut photoexcited with 532 nm laser light | Fungal growth stimulated under laser interference gratings. | 38 |
| HeLa cells | Fe-doped LiNbO$_3$ | x-cut photoexcited with 436 and | Fast necrotic cell death; blue light more | 67 |

| | | 546 nm light (HP Hg lamp); also microparticles were tested | efficient; microparticles induced phototoxicity. | |
|---|---|---|---|---|
| *E. coli* bacteria | $LiNbO_3$ and $LiTaO_3$ | Pyroelectric effect (micro- and NPs) | Bactericidal action under thermal cycling. | 71 |
| Human adipose stem cells | PVDF | Corona domain poling | Enhanced cell attachment; adequate osteogenic differentiation. | 58 |
| HeLa cells | Fe-doped $LiNbO_3$ | NPs photoexcited with green LED light | Cell death under light; however NPs proved cytotoxic by themselves. | 39 |
| MC3T3 osteoblasts | $LiNbO_3$ | z-cut or x-cut mono-domain crystals | Enhanced cell growth and metabolic activity on the z-cut crystal. | 59 |
| Rat bone marrow stem cells | $LiNbO_3$ | z-cut or x-cut mono-domain crystals | Enhanced phenotypic differentiation towards osteoblasts in (+) z LN. | 60 |
| NIH 3T3 mouse fibroblasts | $LiNbO_3$ | z-cut mono-domain crystal | Better cell spread on (-) z; better wound healing on (+) z face. | 29 |
| Saos-2 osteoblasts | $LiNbO_3$ | z-cut mono-domain crystal | Increased cell proliferation, spreading, protein expression ($\beta$1-integrin, vinculin and alkaline phosphatase). | 61 |
| Bone marrow stem cells | PVDF | Corona domain poling | Increased cell proliferation and osteogenic | 62 |

| | | | cell differentiation. | |
|---|---|---|---|---|
| Rat bone marrow stem cells | BaTiO$_3$ NPs | NPs functionalized organic fibers | Increased cell proliferation and osteogenesis stimulation. | 63 |
| Rat bone regeneration | BiFeO$_3$ | (+) and (-) domain nanofilms | Enhanced bone regeneration. | 64 |
| Fibronectin and MC3T3-E1 (mouse osteoblast precursor) | PVDF-TrFE | Contact electric poling | Increased cell proliferation and osteogenic differentiation for moderate surface charges. | 65 |
| Mouse embryonic fibroblasts | Diverse titanates | Spontaneous micro-domains | Enhanced cell proliferation for lead-free titanates. | 66 |
| Chinese hamster lung cells DC-3F | Tourmaline microparticles | Pyroelectric effect | Enhanced bleomycin cytotoxicity presumably due to cell membrane poration. | 72 |

## IV. BIOLOGICAL ACTION MECHANISMS

In this section tentative action mechanisms that can be taking place when biological materials (biomolecules, cells) are exposed to ferroelectrics will be presented. In the spirit of a focused review, the mechanisms will not be discussed in depth, just general arguments and comments will be introduced along with relevant bibliographic references that we think might help clarify future planning, research and modelling in this nascent area. It is important to remark that the following mechanisms are not necessarily mutually exclusive; on the contrary, it is our assumption that most, or all, of them can be occurring at the same time, or in a tight temporal sequence. Therefore, they must be seen as complementary phenomena that can explain the biological responses to ferroelectric materials.

### A. Electrochemistry

The first, obvious, presumable action mechanism one can think of is ferroelectric-driven electrochemistry. Given the large values of the electric field mentioned in previous sections ($10^4$-$10^5$ V cm$^{-1}$) it seems that some kind of electrochemistry must be taken place at the ferroelectric´s surface. Indeed, there are many reports of electrochemical reactions happening with excited ferroelectrics. For example, Tiwari and Dunn described how a photoexcited ferroelectric (e.g. BT or PZT) can reduce $Ag^+$ to metallic $Ag^0$ on the ferroelectric, with a counter oxidation reaction involving $Pb^{2+} \rightarrow Pb^{4+}$ .[73] They also discussed the possibility to photo-electrolyze water to $H_2$ and $O_2$. However, these approaches require UV light (to directly photoexcite the ferroelectric, not any dopant, pumping electrons from the valence band to the conduction band), the use of sacrificial compounds providing electrons, and/or the functionalization of the ferroelectric surface with some better conductor, like a noble metal (e.g. palladium). Similar results or approaches have been reported both for photoexcitation[74,75] and thermal excitation (PY effect).[76-79] Although the experiments show beyond doubt that it is possible to split water and generate ROS in the medium due to excitation of a ferroelectric, the efficiency of the process is particularly low. In fact, ROS concentrations (mainly measured as the presence of the hydroxyl $\cdot OH$ radical) are in the nM to $\mu M$ range. These values are far too low to induce a damaging action on cells, as proposed by Gutmann $et\ al.$,[71] and probably even for physiological redox signaling.[80,81] Previous results from our laboratory, although preliminary, could not find $H_2O_2$ concentrations above 1-5 $\mu M$ (our detection threshold).[67] This would put ferroelectric generated ROS, if produced at all, in the physiological signaling range, perhaps involved in cell attachment or migration, but far below the concentration needed to induce cell damage.

From the published results and modelling, it seems that the direct electrolysis of water on ferroelectric substrates is inefficient in most cases. This seems to be a consequence of the electrochemical kinetic limitation for charge transfer across interfaces between ferroelectrics (in general, poor electric conductors) and the external medium, and the inhibiting role of the double layer established at the surfaces.[73,74] Better results are obtained when there is some kind of catalyst, like a metal, that favors charge movement across interfaces. As such, we consider that ROS generation in excited ferroelectrics is a minor actor in the induction of biological responses. This issue must be studied in further detail, nonetheless, as it could fine tune other major mechanisms that will be treated below.

An alternative electrochemical mechanism deserves a comment. It has been advanced that it can be feasible to generate local regions with pH values different from the average pH of the medium through ferroelectric excitation.[82] These pH changes could signal cells to attach/detach depending on local $H^+/OH^-$ concentrations. Whatever the pH variations are the result of an electrochemical reaction (unlikely) or a local EP process (likely) is an interesting topic to further study. If such pH alterations indeed occur at ferroelectric surfaces, they could have also interest in 2D and 3D isoelectric molecule analysis.[8]

## B. Electrokinesis

Electrokinetic, or electrohydrodynamic, phenomena have a much more relevant role, in our opinion, when it comes to explain the observed results presented in Section III. This is because electric field gradient-driven molecular flows and displacements are established phenomena actuating biological responses, while electrochemistry is not strongly supported as a driving mechanism in experiments with ferroelectrics. For electric fields to induce a biological response two events must take place: first, the electric field must be detected (it must reach or surpass the biological detection threshold), and second, there must be transducing mechanisms that translate the detected electric signal into a biochemical activation sequence.[83] It is important to stress that most examples of electrokinetic detection/stimulation to be discussed here allude to electric field ranges of $10-1000$ mV mm$^{-1}$ ($0.1-10$ V cm$^{-1}$ / $10-1000$ V m$^{-1}$).[6,83] These are far weaker than field strengths customarily induced with ferroelectrics (kV cm$^{-1}$ and higher). But under biological culture conditions the presence of water and dissolved ions surely screen very quickly the surface charges and, hence, the field. In consequence, it is reasonable to accept, at least as a starting point, that field values of the order of physiological ones are more the rule rather than the exception when ferroelectrics interact with cells and biomolecules.

Several biophysical or biochemical transducing mechanisms are currently under scrutiny in order to understand how cells sense and respond to electric fields.[84] To cite a few, the modulation of ionic flows (mainly Ca$^{2+}$) into the cell, changes in the plasma membrane polarization state, alterations in the distribution of membrane proteins, modifications of the cytoskeleton, or biomolecule EP have been suggested or proven to mediate biological responses as varied as cell migration, differentiation, proliferation, or apoptosis.[85] In what follows we will briefly introduce some of these mechanisms, to put them in perspective regarding the data presented in Section III. This discussion, as can be understood, is far from being comprehensive; our aim is to point research directions which we consider interesting for biological applications of ferroelectrics.

One of the most studied mechanisms of action of electric fields is the modulation of ion flows into and out the cell. In particular, Ca$^{2+}$ has received most of the attention, as it is a key cation, not only involved in the overall electrical charge of the cell, but also as a very relevant second messenger in cell signaling (e.g. calmodulin pathway). Under physiological circumstances Ca$^{2+}$ is more concentrated outside the cell. It has been argued that Ca$^{2+}$ can enter the cell under an external electric field by passive flow, to initiate changes in cell migration.[86] These flows would be a consequence of the plasma membrane hyper- or depolarization in response to the imposed external field. In cell regions where a surplus of negative charge accumulates in the inside, Ca$^{2+}$ inflows can take place to compensate the charge. It is known that local changes in [Ca$^{2+}$] lead to cytoskeletal rearrangements, either promoting (low [Ca$^{2+}$]) or inhibiting (high [Ca$^{2+}$]) cell migration.[85,86]

More probable is the inflow of Ca$^{2+}$ through voltage-gated calcium channels when a certain depolarization level is achieved. Voltage-gated, in conjunction with ligand-gated Ca$^{2+}$ channels, are the canonical pathways starting Ca$^{2+}$ signaling in cells. Several authors propose Ca$^{2+}$ signaling as one of the major mechanisms for electric fields to alter cell states, in particular in relation to cell migration, attachment, axonal growth, and wound healing.[83,85-89] Therefore, intracellular [Ca$^{2+}$] changes seem to us a very important signaling element, that may have a crucial role in the observed cell responses to excited ferroelectrics.[28-31,35,59] As Ca$^{2+}$ has also been involved in the regulation of other critical responses such as cell proliferation,[85] contractibility,[86] or differentiation/dedifferentiation[88] to name a few, this signaling ion could have a relevant role in other reported biological responses to ferroelectrics: osteoblastic differentiation[60,64] or enhanced proliferation.[61-66] It deserves mentioning that, although Ca$^{2+}$ signaling features the largest number of studies in the field of electric biomodulation, there are some works shedding light upon other relevant voltage-gated channels, like Na$^+$ channels[86] or K$^+$ channels,[85] to explain cellular responses to electric fields.

A complementary mechanism, with a very relevant role, is the electric field-driven displacement of large biomolecules: mainly proteins, protein complexes, and other charged biopolymers (e.g. proteoglycans/glycosaminoglycans). This displacement is the result of EP and/or electro-osmosis. That EP can spatially polarize charged biomolecules comes as no surprise. However, some results obtained in the 1970s and early 1980s showed that certain membrane proteins and channels, expected to migrate to one pole of the cell due to their intrinsic charge, localized at the opposite cell side under an electric field. The phenomenon of electro-osmosis was invoked to explain these paradoxical results.[90] Electro-osmosis involves the movement of water (or other non-charged fluids) coupled to the EP of some charged entity, like dissolved ions, for example (see Fig. 10 (a)). Under the correct conditions of electric field strength, ion concentration, overall charge, temperature, pressure, etc. it is feasible for the electro-osmotic flow to drive charged molecules "in reverse", if we attended only to EP (see Fig. 10 (b)). Electro-osmosis is capable of concentrating membrane proteins at certain regions or poles of the cell. Once an asymmetric biomolecule distribution is established, it can drive biological responses to the electric field due to the concomitant localized metabolic activity (see Fig. 10 (c)).[83,85,87,88]

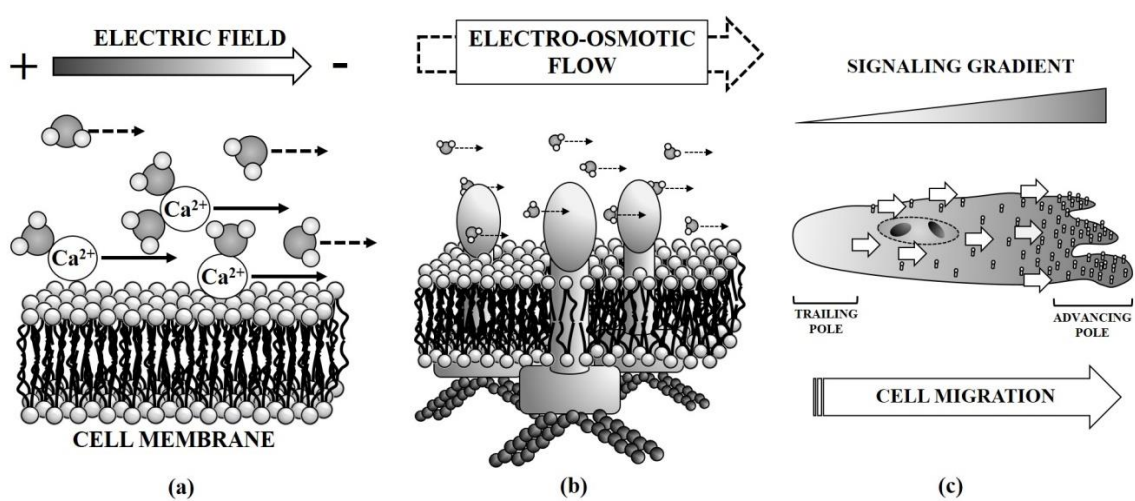

FIG. 10. Scheme of the electro-osmotic process that can take place at the plasma membrane when a cell is exposed to an electric field. (a) Under the influence of an external electric field ions ($Ca^{2+}$ show as an example) migrate due to EP (black arrows), some close to the cell membrane. They drag along their hydration shell (water molecules "attached" to $Ca^{2+}$). Free water molecules can be entrained by this flow (electro-osmotic flow) if strong enough, and move in the same general direction (dashed arrows). (b) The electro-osmotic flow can induce displacement of membrane structures, like proteins (transmembrane structures), due to molecular drag (dashed arrows). These structures can relay signals to the cell or, as shown in the cartoon, they can be attached to the cytoskeleton (dark strings beneath the proteins). (c) Sustained electro-osmosis (white arrows) can polarize the cell membrane over time (proteins move and accumulate at the right cell pole). This can lead to enhanced or inhibited signaling (signaling gradient) that can drive, for example, cell migration in certain directions. Structures not to scale in the schemes.

Two examples of this functional membrane polarization will help clarify the mechanisms. Recently, Huang *et al.* published results in which heparin sulfate (an abundant proteoglycan present at the cell surface) polarizes in respond to fields of the order of 1 V $cm^{-1}$.[91] In several neuronal-related cell types they found that this polarization invariably led to cell migration. The molecular action mechanism proposed relates the increased accumulation of surface heparin sulfate on the outside to the internal inhibition of Rho GTPases Rac1 and Cdc42. Similar results were recently published by another group In their experiments and model the cell membrane lipid rafts are the structures strongly polarized under the imposed electric field.[92] Lipid rafts are multiprotein structures located at the cell membrane which mediate many cues from the external environment to the cell interior. Among other functions, they are crucial for the membrane-cytoskeleton coordination, which of course is primordial for cell migration and attachment. Under fields of 6 V cm$^{-1}$ lipid rafts accumulated at one side of the cell, and initiated signaling cascades involving integrins, caveolins, RhoA, Src kinase and PI3 kinase. All these proteins are very relevant in many cellular processes. Additionally, these authors found very interesting results also involving polarized lipid rafts when the cell cultures were exposed to low-frequency AC fields. In connection to all of these, Vaněk *et al.* found increased cell spreading, attachment, and higher levels of β1-integrin and vinculin in osteoblasts exposed to z-cut LN (see Section. III.B.1).[61] As Rac1 mediates cell galvanotaxis (100 mV mm$^{-1}$) responses activated by the membrane proteins integrin β4 and epidermal growth factor (EGF) in human keratinocytes,[93] this protein should be a target for further study in regards to biological responses to ferroelectrics. A whole set of signal transduction, mechanotransduction and osteogenic differentiation protein expression is provided by Liu *et al.* in their BFO-driven bone regeneration experiments.[64] In summary, it can be highly enlightening to assess the membrane distribution of lipid rafts or proteoglycans in cells exposed to ferroelectrics. Also, elevated intracellular levels of ROS (perhaps a ROS gradient) should be assessed under these circumstances (see Section IV.D below), as activation of the Rac family of GTPases is known to induce activation of NADPH Oxidases (NOX) and downstream ROS signaling.[85,89,94-96]

Given space constrains, we direct the reader to the excellent and very recent review by Thrivikraman *et al.* for additional information and sources on the cellular mechanisms of action of weak electric fields.[85] Other mechanisms proposed and discussed in several publications include action of polarized transmembrane proteins on the cytoskeleton,[92] intracellular protein EP,[87,88] protein/transcription factor voltage-dependent activity,[88,89] and role of heat shock proteins.[85]

## C. Molecular crowding

Molecular crowding, generally speaking, occurs when a substantial fraction of the volume in a solution is taken up by solute molecules instead of solvent molecules.[97,98] This is exactly the situation in biological milieu: around 30-40% of the cell volume is made of bio- and macromolecules (proteins, lipids, nucleic acids, carbohydrates, cofactors, etc.).[99,100] Under this circumstance, steric factors become very important and chemical kinetics are strongly modified as compared to dilute concentrations. The system behavior becomes non-linear depending on the solute concentration. Molecular crowding has a decisive impact on biological function and regulation (see also Section IV.D below), and several publications highlight its importance to replicate physiological-like conditions to obtain optimized biological responses (regeneration, tissue culture).[101-103]

Electric fields generated by ferroelectrics can presumably induce electrokinetic processes, as discussed in Section IV.B.2 (see above). These electrokinetic processes (EP, DEP, or electro-osmosis, for example) may concentrate biomolecules in certain surface locations, thus mimicking a molecular crowding effect at the local (micrometric) level.

This, in turn, may stimulate a variety of cellular responses like enhancement of cell attachment, spreading or extracellular matrix deposition.[101] Non-uniform electric field arrangements lead to electro-osmotic flow inception in ionic solutions.[104] These flows spontaneously generate vortices close to surfaces, which invariably favor mass transport and deposition in certain regions. Should this be happening with ferroelectrics, these accumulations could drive biomolecule concentration and cell interactions with these "privileged" spots. In this sense, Zhang *et al.* reported that d-cysteine strongly accumulates at (+) domains on the surface of PPLN,[105] and Haußmann *et al.* describe the nanoscale photodeposition of two organic fluorophores (rhodamine 6G and Alexa fluor 647) over polarized domains of LN.[106] And, in a more biologically-oriented paper, another group recently published that positively-charged tertiary amines functionalizing polyethylene surfaces enhanced osteogenesis of rat bone marrow mesenchymal stem cells as compared with non-functionalized surfaces.[107] The important positive surface charge imposed by the amines induced a robust upregulation of inducible nitric oxide synthase (iNOS) in the attached cells, which drove downstream transcription of osteogenic genes like alkaline phosphatase, Runx-2, OCN and BSP. Very similar results have been published even more recently, in which biocompatible positively-charged polymers promoted osteogenic differentiation of mesenchymal stromal cells.[108] At least six genetic markers of osteogenesis (among them Runx-2, OCN and BSP) were upregulated, as well as a robust alizarin red test response. The authors advance that the observed responses are the result of activation of the ephrinB2-EphB4 signaling pathway. All of this reminds of the results obtained with ferroelectrics by several groups (see Section III.B),[59,61,62,64,65] and seems particularly relevant in regards to the results reported by three groups as to the osteoblastic differentiation of bone marrow stem cells on positive ferroelectric surfaces.[60,63,64] Hence, as advanced by several papers, a preferred ionic and molecular concentration (ionic/molecular crowding) at ferroelectric surfaces seems a very logical and attractive hypothesis to explain the results observed so far.[29,33,34,50,53,59]

The surface polarization can, additionally, favor rheological changes at very short distances, like increases in viscosity[109] or water structuring.[101] Arguably, these changes can arise as the result of the high surface charge in combination with the concentrated ionic medium, which, in turn, would promote the establishment of a wide double layer and Stern layer. These may signal the cell to increase or decrease interactions with said surfaces. It is convenient to remind at this point that both the extracellular matrix and the cell membrane, along its immediate surroundings, are localizations with important water structuring due to the charged macromolecules (proteoglycans, proteins) present.[97-99] These environments are far from diluted solutions, where random thermal movement is the norm. Ferroelectrics, then, could well be mimicking a "surrogate" extracellular matrix to some extent, "inviting" the cell to respond accordingly.[101] In this line, the results reported by Ribeiro *et al.* highlighted an electrically-driven accumulation of FN over poled-PVDF surfaces, although the cell impact of this accumulation was unclear.[27] Then, further work in this line pointed to the role of electrically-increased surface hydrophilicity to provide the correct environment for biomolecule-cell interaction to take place.[58,62] In-depth results and discussion in this sense have been provided by Tang *et al.*[65] Their very interesting results and computer simulations point to a double effect related to molecular adsorption and crowding. On the one hand, biomolecules (fibronectin in their paper) show a trend of increased accumulation on a ferroelectric surface as the surface charge increases. At the same time, the biomolecules display steric effects, probably driven by electrostatics but arguably also by crowding, that enhance the efficiency of their interaction with cells. In their experiments this enhancement reaches a saturation, after which cell interaction diminishes. Thus, a caution remark seems appropriate as to the positive vs. negative effect of too much surface charging in regards to the biological outcome.

## D. Electroporation and cell volume signaling

Finally, we will briefly discuss the possibility that cells exposed to ferroelectrics undergo an electroporation process which can result in biological modulation (if mild) or cell damage and death (if strong). Electroporation is the process in virtue of which the cell membrane is compromised and loses its integrity when exposed to a strong electric field.[110] The integrity loss occurs as a result of the creation of membrane pores that allow the spontaneous movement of ions and molecules (including water) down their particular electrochemical gradients. Electroporation models assume that nanometric (1-2 nm) pores are spontaneous and continually formed in the cell membrane as a consequence of thermal noise.[111] As soon as they form, however, they seal due to hydrophobic forces and there is no net (or very small) transfer across the cell membrane. In the presence of an electric field, and given that biological membranes are made of charged phospholipids, these "virtual" pores can be influenced by the field and grow beyond the diameter where they spontaneously reseal.[112,113] Under these conditions unimpeded flow across the membrane takes place, with important consequences for the cell metabolism. Water will enter the cell under osmotic gradients and cell swelling will ensue. If cell membrane compromise is widespread and lasts long enough, the cell will undergo osmotic shock, swelling and necrotic death due to excessive internal volume and biomolecule dilution.[5]

Our results from 2011 point precisely in this direction, as our treatment led to a cell morphology fully compatible with a massive osmotic shock, developed in a matter of minutes during light exposure (see Fig. 9).[67] The early cell swelling was followed by large bubbles evolution, clear signs of a severely compromised cell membrane integrity.[68,69] As mentioned previously, we failed to detect any $H_2O_2$ that could explain membrane damage due to oxidative chemistry. On the other hand, very high electric fields are achievable and sustainable under light intensities similar to the ones employed in our experiments ($10^3$-$10^4$ V cm$^{-1}$).[114] Such electric fields and exposure times (minutes) are adequate to induce cell death by electroporation.[5]

The possibility to induce cell electroporation through ferroelectric excitation opens other interesting avenues that we are starting to explore. If milder conditions can be produced (shorter exposures, weaker fields, etc.), it is plausible that better control over the electroporation process can be achieved. Under this scenario cells can be forced to uptake molecules or compounds to which the membrane is usually impervious. This is precisely in line with the results reported by García-Sánchez *et al*. (Section III.B.3) where bleomycin uptake and cytotoxic action was promoted by the PY effect of tourmaline.[72] In line with this, this methodology could be employed to gently swell cells and recover metabolites of interest, for example, for analytical purposes. Furthermore, electroporation could be synergistically enhancing the bactericidal ROS action observed when pyroelectric ferroelectrics were thermally cycled.[71]

Beyond that, we envision a certain cell control through manipulation of the cell volume.[115] As it turns out, cell volume is one of the key metabolic regulators of the cell. Animal cells maintain a delicate osmotic balance which provides the adequate cell volume.

For example, it is obvious that a proliferating cell must roughly double its initial volume if it is to divide and produce two daughter cells. Hence, such vital processes as proliferation, migration, or apoptosis depend on a tightly regulated cell volume control to proceed.[116-119] Through adequate ferroelectric interaction a mild electroporation process can change the volume of cells and influence in their physiologies. Such a process could lead to a gentle cell volume increase, enough to signal through membrane stretching, integrins, Src kinases and Rac (Rho GTPases) to induce a manifold of biological responses,[120-124] as well as modify the internal molecular crowding state.[125] These processes, we believe, are very important in relation to cell migration and wound healing (see Sections III.B and IV.B). Weak electric fields can lead to a physiological "electroporation", not by disrupting the phospholipid integrity, but by facilitating membrane channels opening (or remain in the open state for a longer time). This phenomenon also engages cell metabolism through changes in cell volume.[126] For example, $Ca^{2+}$ gradients and waves are established and sustained during cell migration, also under the influence of electric fields (see also Section IV.B).[127] Cell remodeling can be seen as an increase in cell volume at the leading edge and a concomitant volume decrease at the trailing edge. Electric fields induced by ferroelectrics can enhance these volume changes by interfering with the ion channels, or directly producing a very small population of membrane pores by a gentle electroporation.

We have summarized what we consider the most plausible action mechanisms to explain each published reported included in this focused review. Table III features the publications along with the biological responses reported, excitation approaches and plausible mechanisms at work. As remarked previously, they should not be considered as exclusive one of another. On the contrary, it is our working hypothesis that these phenomena are taking place more or less concurrently, and that they are, to a great extent, interlocked. For example, electro-osmotic displacement of membrane proteins and /or lipid rafts may signal by itself, but also can make certain regions of the membrane more or less sensitive to electroporation (e.g. by locally altering membrane fluidity when biomolecule composition/ratios are changed). Therefore, caution is necessary when biological responses are discussed in the framework of ferroelectrics. One of our aims is, precisely, to warn about the necessity to consider several biophysical processes when the time comes to plan the experiments and discuss the results.

TABLE III. Biophysical mechanisms tentatively producing the reported biological responses to ferroelectric materials.

| Biological response | Ferroelectric induction mechanism | Biophysical mechanisms | References |
|---|---|---|---|
| Biological structures and molecules trapping | Polarized domains; PV | EP; DEP; Molecular crowding | 26, 27, 39, 40, 49, 65 |
| Cell trapping | Polarized domains; PV; PY | EP; DEP; Electrokinesis; Molecular crowding | 28, 38, 47, 22, 51, 52, 53 |
| Cell attachment, spreading and | Polarized domains; Polarized surfaces; | DEP; Electrokinesis; Molecular crowding; | 27, 28, 29, 30, 31,35, 38, 58, 59, 60, 61, |

| growth | Ferroelectric NPs; PV | | 62, 63, 64, 65, 66 |
|---|---|---|---|
| Directed cell growth | Polarized domains; Polarized surfaces; PV | DEP; Electrokinesis; Molecular crowding | 35, 47 |
| Cell migration and wound healing | Polarized surfaces; | Molecular crowding | 29, 65 |
| Cell differentiation | Polarized surfaces; Ferroelectric NPs | Electrokinesis; Molecular crowding; | 58, 59, 60, 61, 62, 63, 64, 65 |
| Membrane poration and molecule uptake | PY | Electroporation | 72 |
| Membrane poration and cell death | PV; PY | Electrochemistry; Electroporation | 39, 67, 71 |
| In vivo bone regeneration | Polarized surfaces | EP; DEP; Electrokinesis; Molecular crowding | 64 |

EP: electrophoresis; DEP: dielectrophoresis; PV: bulk photovoltaic effect; PY: pyroelectric effect.

# V. CONCLUSIONS AND PERSPECTIVES

The study and use of ferroelectric materials in biophysics, biology and biomedicine is quite a novel scientific endeavor. As such, most questions remain unanswered and most action mechanisms are to be defined and measured. But it is precisely these fertile grounds, we believe, that will provide the high appeal and strong momentum for researchers at the overlap of many scientific areas to study these biological applications. The featured early experimental results and tentative mechanistic models set the stage for a vigorous expansion of the field. Several advantageous characteristics make these materials attractive as a new, radical approach for bioelectricity research: the avoidance of an external electric source; the possibility to work with semi-permanent "electrode regions" (polarized domains) spanning several orders of magnitude in size (from cm down to micrometric); alternatively, extreme flexibility to generate "virtual electrodes" *ad hoc* by taking advantage of the PV or PY effects; resilient and biocompatible materials; and foreseeable integration into microfluidic devices.

So far, only a few ferroelectric have been used in the biological field: notably $LiNbO_3$ (pure and Fe-doped), but also $LiTaO_3$, $BatiO_3$ and some other titanates, and organic ferroelectric PVDF films. Mono-domain crystals, domain structures and even micro and nanoparticles have found a variety of applications, as discussed above. In this sense, the biotechnology field can, and should, take advantage of the important research activity and recent progress in ferroelectric domain engineering. Additionally, for applications involving nanoparticles, a key point that still deserves much attention is the investigation of the ferroelectric (photovoltaic or pyroelectric) properties at the nanoscale.

Employing these materials it has been shown that electronic tweezers are feasible, and cell trapping has been proved, although more systematic studies are necessary. For the moment, trapping has been reported in 2D. But it should be possible to attain full 3D manipulation my merging EP/DEP and microfluidic forces, or by making use of more than one ferroelectric surface (e.g. ferroelectric-lined channels). Several groups have proved that patterning of biological structures is robust down to microns. This patterning can lead to interesting possibilities, for example, antibody "coating" into predetermined areas or patterns, or even reconfigurable ones making use of virtual electrodes.

It is of utmost importance to define in depth the biophysical and biological mechanisms that are actually taking place when cells interact with ferroelectrics. A first attempt to advance some tentative mechanisms has been provided in Section IV. But the experimental proof supporting these mechanisms is, at present, indirect and feeble. Therefore, a whole range of new studies should focus on defining and provide direct proof of which mechanisms are really at work. In our opinion, electrokinetically-driven flows (EP/DEP and electro-osmosis) are very interesting candidates to partly explain the biological action observed so far. In experiments with a strong ferroelectric excitation (intense light excitation of the PV effect) cell membrane electroporation seems a very reasonable action mechanism. Also, it is very important to definitively settle down or provide support for the possibility of electrochemistry taking place. If water, or solutes, are or not being electrolyzed is an important subject to determine. Future applications would most probably differ if electrochemistry is taking place or not at the ferroelectric surface. As a corollary to this, the electrochemical production of ROS in these systems needs to be carefully addressed. These chemical compounds can be very damaging but, on the other hand, can physiologically signal at low concentrations for short exposures.

The analytical possibilities seem very large. A full range of conditions to study cell metabolites and components can be envisioned. From a dramatic cell bursting, where all cell contents are spilled into the medium with total destruction of the cell membrane, to a gentle cell swelling to study ionic and osmolite flows under an imposed cell volume increase that keeps the cell alive. These materials can offer a new approach to study cell volume regulation. For example, selected cells can be studied and challenged within a cell population by making use of virtual electrodes, while manipulating the ionic composition of the medium.

Studies focusing on galvanotaxis, cell migration, wound healing, tissue regeneration, and cell proliferation await testing and assessment making use of ferroelectrics. Cells can be analyzed after punctual excitation of the ferroelectric or during continuous field generation. Although our preliminary impressions on antitumoral applications are precautious (given the size of the substrate needed to obtain strong enough fields), it nevertheless deserves the effort of studying, as there can be experimental conditions under which the outcome is positive. In this, and other possible applications, it is necessary to better evaluate the biological responses obtained with micro- and nanoparticles of different ferroelectrics.

In conclusion, it is our believe that the field of biological applications of ferroelectric materials offers a broad new horizon of scientific possibilities to exploit at present, both theoretical and experimental. We aim to stimulate and stir the scientific community, both already involved in this field as well as new groups, into a deeper and broader study of these issues. We have tried to provide, not only a descriptive revision of the published results, but also some reasonable action mechanisms that help researchers to find a starting point, from which to plan experiments and think of plausible biophysical phenomena explaining the obtained data. We hope that this focused review will provide the needed momentum to expand the study of biological applications of ferroelectrics on a larger scale.

**ACKOWLEDGMENTS**

We are grateful to the reviewer for suggesting relevant additional references. Funding from the Ministerio de Ciencia, Innovación y Universidades under projects MAT2014-57704-C3-1-R and MAT2017-83951-R is acknowledge. A. Blázquez-Castro acknowledges funding under the Marie Skłodowska-Curie Action COFUND 2015 (EU project 713366 – InterTalentum), and partial financial support under EMBO STF 7527 fellowship.